\documentclass[conference]{IEEEtran}

\ifCLASSINFOpdf

\else

\fi

\usepackage{placeins}
\usepackage{graphicx} 
\usepackage{amsmath,amsfonts}
\usepackage{mathtools}   
\usepackage{bm}          
\usepackage{siunitx}     
\usepackage{algorithm}          
\usepackage{algpseudocode}      
\usepackage{booktabs}

\usepackage[
  backend=biber,
  style=numeric,
  sorting=none,
  doi=false,
  url=false,
  isbn=false,
  eprint=false
]{biblatex}

\AtEveryBibitem{%
  \clearfield{url}%
  \clearfield{urldate}%
  \clearfield{eprint}%
  \clearfield{note}%
}

\newcommand{\quotes}[1]{``#1''} 

\begin{document}

\title{{Crypto-Microeconomics: The Distribution of Bitcoin Wealth Among Diverse Economic Agents}}
\author{\IEEEauthorblockN{Syed Azhar Hussain}
\IEEEauthorblockA{Munster Technological University, \\Ireland\\
}
\and
\IEEEauthorblockN{Kashif Ahmad}
\IEEEauthorblockA{Munster Technological University, \\Ireland\\
}
\and
\IEEEauthorblockN{Mubashir Husain Rehmani}
\IEEEauthorblockA{Munster Technological University, \\Ireland
}}

\maketitle 

\begin{abstract}
Bitcoin (BTC) wealth distribution is often analyzed with macro indicators such as aggregate addresses, wallet balances, prices, network activity, fees, and hashrate. This letter applies a Crypto-Microeconomic perspective to examine wealth concentration across five labeled economic-agent classes: Service, Abuse, Malware, Individuals, and Benign. Using descriptive statistics, inequality metrics, and longitudinal indicators, we show that Bitcoin wealth is highly concentrated across major classes. Service entities hold the largest share of observed BTC (75.15\%), while Abuse controls a disproportionately large share relative to its entity count (24.26\% of BTC vs. 3.53\% of entities). Individuals, Abuse, and Service show near-maximal within-class inequality (e.g., Gini \(=0.9993\) for Individuals), and time-series analysis indicates these patterns still exist. Overall, Bitcoin wealth among labeled economic agents remains structurally uneven and concentrated in a small subset of entities.
\end{abstract}


\IEEEpeerreviewmaketitle

\section{Introduction}

Macroeconomic evaluations employ aggregate data to track wealth distribution patterns \cite{sai_characterizing_2021}, while microeconomic models examine how the behavior and market participation of individual economic agents influence income and pay distribution \cite{VENKATASUBRAMANIAN2015120} as well as Bitcoin's supply, demand, and value \cite{halaburdaMicroeconomicsCryptocurrencies2022}, thereby potentially contributing to wealth inequality. In both conventional and crypto-currency economies, analyses at both macro- and microlevels identify the effects of capital accumulation and disparities in wealth. In Bitcoin, understanding the wealth distribution is essential for assessing claims about decentralization. In addition, it contributes to strengthening security by uncovering vulnerabilities associated with wealth aggregation and offers deeper economic insights into the behavior and stability of the evolving crypto-economy \cite{ilinskii_wealth_2019}. Initially promoted as an instrument to democratize finance and address economic disparity, Bitcoin's wealth allocation frequently mirrors or worsens conventional economic inequalities \cite{sai_characterizing_2021}, and is widely regarded as highly volatile and speculative asset \cite{smales2019bitcoin}, with limited adoption as a medium of exchange, even where recognized as legal tender \cite{ElSalvadortendor}.

In the crypto-economy, complex networks connected with deanonymized Bitcoin holders, wallets, and addresses \cite{ilinskii_wealth_2019} have presented challenges to the investigation of wealth distribution. Traditionally, the examination of the wealth distribution within the Bitcoin ecosystem has utilized \quotes{macroeconomic indicators}, including wallet and account balances, market prices, mining pools attributes, network activity, among others, analyzed with widely recognized wealth inequality metrics. Also recent literature about crypto-microeconomics by \cite{halaburdaMicroeconomicsCryptocurrencies2022} examines the supply, demand, market value, and competition of cryptocurrencies, focusing on the incentives of miners, users, exchanges, and rival networks. Moreover, these empirical investigations rely on established wealth inequality metrics such as Lorentz and Lame curves, the Herfindahl-Hirschman index, Pareto, Zipf's laws, Top-k wealth shares, Shannon entropy, Gini coefficient, and Nakamoto coefficient \cite{cowell2011measuring, juodis_overview_2024, ilinskii_wealth_2019, de_marzo_growing_2023}. 

Specifically, the Gini coefficient measures wealth inequality, while the Nakamoto index evaluates wealth concentration at a threshold 51\%, particularly relevant in Proof-of-Work cryptocurrencies \cite{sai_characterizing_2021}. The macroeconomic assessment of Bitcoin revealed that its Gini coefficient decreased from an extreme 0.997 in 2013 to a still high 0.653 in 2023 \cite{juodis_overview_2024}, opposing cryptoanarchy ideals. Unlike traditional microeconomic research, which provides valuable insight into the factors influencing wealth disparity and its economic impacts, crypto-microeconomic studies are seldom explored. The major gaps in Bitcoin wealth micro-analysis include the following:
\begin{itemize}
    \item Existing analytical research is largely based on the Bitcoin Rich List based on Top-k wealth shares and investigate wallets and associated addresses. This method is inadequate for analyzing crypto-microeconomics because it overlooks complex, often obscure interactions among market agents, including individuals, wallets, addresses \cite{ilinskii_wealth_2019, sai_characterizing_2021}, and market segments (groups).
    \item The application of Lorentz and Lame curves along with the Gini coefficient facilitates macro-level insight into Bitcoin wealth distribution, although they remain unexamined in addressing microeconomic imbalances among various market segments \cite{ilinskii_wealth_2019} i.e., classes of economic agents.
    \item Studying microeconomic inequality requires identifying individual economic agent classes in the Bitcoin ecosystem, where pseudonymity complicates it, however improved deanonymization of market participants can mitigate the problem \cite{sai_characterizing_2021, ilinskii_wealth_2019, mizerkaRoleBitcoinDeveloped2020}.
\end{itemize}

The main contribution of this study is to address significant gaps in the literature on crypto-economics by providing a structured microeconomic examination of the distribution of Bitcoin wealth across five different classes of economic agents. Adopting a agent-class based perspective that contrasts with earlier aggregate analyses, we examine which types of participants are driving the concentration of Bitcoin's wealth, and how the extent of concentration within each class varies across individual actors. Earlier studies has demonstrated a significant association between concentrated cryptocurrency holdings and elevated security and market manipulation risks \cite{sai_characterizing_2021}, whale-driven market dynamics \cite{goutte_cryptocurrencies_2021}, and the influence of large users on Bitcoin returns \cite{mizerkaRoleBitcoinDeveloped2020}. Therefore, our letter examines Bitcoin concentration by agent-class as an indicator of potential market-structure risk: high concentration in Services may reflect aggregation at financial intermediaries, concentration in Abuse and Malware may signify enforcement-sensitive or illicit capital, and concentration in Individuals may represent exposure to large private holders.

The purpose of our study is threefold. First, reconstructs Bitcoin clustered entities into five distinct categories of economic agents: I. \emph{Service Providers} that handle funds (e.g., exchanges, mixers, payment processors, mining operations, tormarkets, gambling platforms, etc.), II. \emph{Abuse Entities} engaged in illicit or criminal activities (e.g., Ponzi schemes, sextortion, theft, terrorism, etc.), III. \emph{Malware Operators} responsible for malicious software linked to ransom collection, cryptojacking, banking trojans, web-skimming, and related attacks, IV. \emph{Individual Users} whose identities can be associated with usernames on forums such as BitcoinTalk and V. \emph{Benign Actors}, including government or donation addresses disclosed publicly. We then measure how observed Bitcoin holdings are distributed both within and across categories of economic agents. Second, we analyzed large-entity wealth distribution within these economic agents by investigating variations in concentration patterns over time using established inequality and concentration metrics. Third, we formulates these concentration patterns as a problem of crypto-microeconomic observability at the agent-class level, demonstrating that overall Bitcoin concentration can be systematically disaggregated into economically meaningful agent categories, rather than being analyzed solely as a uniform, network-wide characteristic.

\section{Background and Related Work}

\subsection{Statistical Evaluations of Bitcoin's Wealth Inequality}
Wealth inequality in Bitcoin is assessed via statistical analysis of coin distribution across addresses, along with conventional economic measures. Macrolevel studies utilize public transaction ledgers to grasp the broader crypto-economy \cite{sai_characterizing_2021}. Researchers draw on secondary data sources, including publicly accessible Bitcoin blockchain datasets, the \quotes{Bitcoin Rich List}, and services such as Bitinfocharts, to examine patterns of Bitcoin wealth concentration and user behavior \cite{goutte_cryptocurrencies_2021, ilinskii_wealth_2019, mizerkaRoleBitcoinDeveloped2020}. They analyze address balances (Unspent Transaction Output - UTXO) and set monetary thresholds for negligible balances to assess wealth distribution \cite{sai_characterizing_2021}. Inequality metrics such as the Gini coefficient reveal strong wealth disparities. The Nakamoto index identifies the smallest group controlling 51\% of a cryptocurrency, linking wealth concentration to security risks \cite{sai_characterizing_2021, juodis_overview_2024}. At the address level, the Herfindahl-Hirschman index (HHI) shows a low value of 19, reflecting significant decentralization \cite{juodis_overview_2024}. 

All these Macro-level Empirical findings frequently reveal substantial wealth concentration in Bitcoin whales. For example, a study found that 0. 01\% of Bitcoin whales addresses controlled almost 42\% of Bitcoin value, with the top 9.7\% owning 99\% of all Bitcoins \cite{goutte_cryptocurrencies_2021}. \cite{mizerkaRoleBitcoinDeveloped2020} show that concentrated user activity shapes Bitcoin price dynamics. Their user graph analysis reveals extreme wealth inequality (Gini Coefficient $\approx\ 1$) and a statistically significant link between major users’ transactions and Bitcoin returns, indicating a possible \quotes{Whale-Effect}. Another study indicated that over 58.21\% of Bitcoins are held by only 0.01\% of large-entities known as \quotes{whale} \cite{sai_characterizing_2021}. This elevated concentration entails the risk of a \quotes{Whale-Effect}, in which these large-entities influence overall market behavior and amplify concentration-related concerns. Most existing studies examine wealth distribution in the crypto economy at a macro level and overlook whale-related risks across different economic agent-classes (micro-level). In contrast, this letter examine which agent-classes serve as primary holders of wealth and quantitatively assess the extent of wealth concentration both within and across these categories.

\subsection{Bitcoin Wealth Analysis Approaches}
Existing literature proposes four-steps method for econometric analysis of cryptocurrencies, involving data source selection, data processing to derive econometric metrics, and comprehensive analysis of individual cryptocurrencies, notably Bitcoin and Ethereum \cite{sai_characterizing_2021}. It uses a Big Data Analytics pipeline with an Extract, Transform, and Load (ETL) process to create a unified repository of Bitcoin's UTXO and Ethereum's balance models into a single \quotes{address-balance model} that includes address, balance, and date details enabling time series analysis of the crypto-economy \cite{sai_characterizing_2021}. Some studies use comparative analysis with third-party data for selected samples to conduct econometric evaluations \cite{ilinskii_wealth_2019, goutte_cryptocurrencies_2021}. However, limited studies use deanonymization methods for granular wealth analysis as proposed here, combining full-ledger data extraction, label-based agent attribution, balance reconstruction, and class-level wealth concentration measurement.
 
Specifically, deanonymization and clustering methods improve Bitcoin wealth inequality analysis by identifying wealth-holding entities. Bitcoin’s pseudo-anonymity and many single-use addresses hinder precise micro-level assessments \cite{sai_characterizing_2021}. Researchers apply reverse engineering heuristics and machine learning to cluster addresses into wallets to enhance de-anonymization. For example, a gradient-boosting classifier deanonymizes the exchanges and merchant services as the main wealth accumulators, increasing the Bitcoin Gini coefficient from 0.65 to 0.73 when the clustering addresses method was used \cite{sai_characterizing_2021}. While these heuristic and back-and-forth analyses reveal transaction patterns and cybercrime structures \cite{gomez_watch_2022}, they neglect granular wealth distribution analysis. Our letter specifically addresses this gap by examining micro-level wealth distribution and the types of key actors in the crypto-economy.

Prior research approches shows that Bitcoin wealth is highly concentrated across addresses, wallets, user-graph structures, and the network as a whole. The key unresolved issue is not the presence of this concentration, but the extent to which it can be systematically linked to economically relevant categories of actors that influence return behavior, market-structure risk, security vulnerabilities, and illicit-payment pathways \cite{sai_characterizing_2021,mizerkaRoleBitcoinDeveloped2020,gomez_watch_2022,Oosthoek2023}. Existing studies offer only limited evidence on how BTC is distributed among Services, Abuse-related actors, Malware-linked entities, Individuals, and Benign users after address clusters are inferred and entities labeled. This paper fills that gap by shifting focus from aggregate, ledger-wide measures to clearly defined, labeled classes of economic agents.

\section{Our Methodology}
\subsection{System Description}
\begin{figure}[htbp]
    \centering
  \includegraphics[
    width=\linewidth,
    height=\textheight,
    keepaspectratio
  ]
    {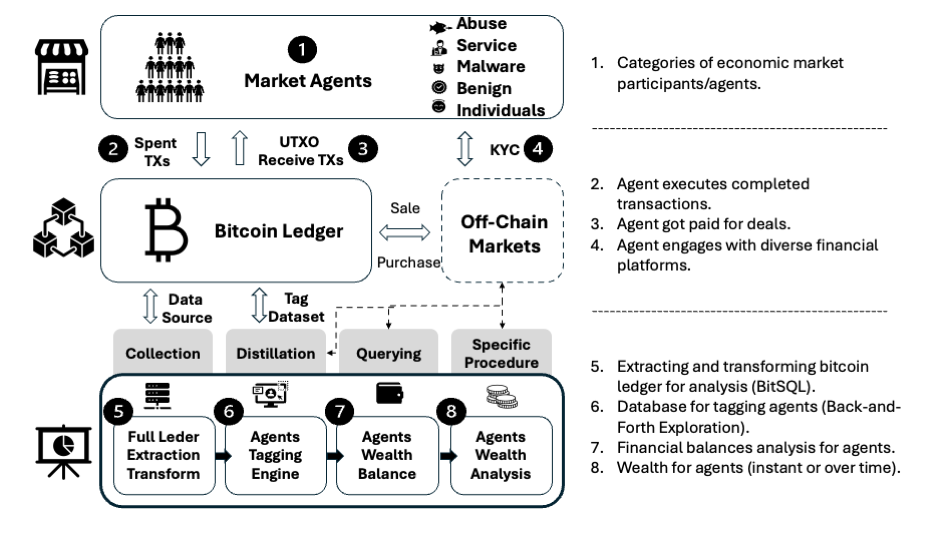}
    \caption{Crypto‑Microeconomic observability method for class‑specific analysis of Bitcoin wealth concentration. The pipeline combines blockchain data extraction, address‑to‑entity clustering, economic‑agent classification, balance reconstruction, and inequality metrics to systematically examine observable BTC holdings across defined agent categories.}
    \label{fig:crypto_micro_wealth_flow}
\end{figure}

We develop a system for analyzing the distribution of wealth among market participants in the crypto-economy over both short- and long-term horizons. As illustrated in Fig. \ref{fig:crypto_micro_wealth_flow}, the Crypto-Microeconomic observability method is implemented as a layered ETL (Extract, Transform and Load) analytical system. The system first ingests Bitcoin ledger data from a local node via BitSQL \cite{mun_bitsql_2022} and then applies entity-clustering techniques \cite{gomez_watch_2022} to aggregate individual addresses into inferred entities (agent-classes). 

These entities are then assigned to five economic-agent classes such as Abuse, Service, Malware, Benign, and Individuals, which enables reconstruction of observed BTC balances. The method also computes agent-class-level concentration metrics, longitudinal concentration measures, and large-entity wealth concentration patterns, evaluated at specific observation points and monthly intervals.

The system study a particular economic phenomenon in which observed BTC holdings are concentrated both within and across labeled economic-agent classes. It also substantially advances the empirical analysis of the large-entity concentration from the Genesis Block event and conducts a detailed examination of the corresponding patterns of wealth inequality across distinct categories of economic agents over the key milestones in Bitcoin’s development.

\subsection{Model Formulation}
We analyze the crypto-microeconomic behavior of economic agents using descriptive statistics. The Top-$K$ share ratio \cite{juodis_overview_2024} measures the percentage of BTC held by the top 1\%, 5\%, and 20\% of positive-balance entities within each agent class, and the Gini coefficient \cite{sai_characterizing_2021} measures intra-class wealth inequality.

\paragraph*{Economic Agents}
Let \(\mathcal{A}=\{\text{Service},\text{Abuse},\text{Malware},\text{Individuals},\text{Benign}\}\) denote the set of clustered agent classes, and let \(\mathcal{E}_a\) be the set of entities in class \(a\). Let \(u\) denote the observation point, where \(u=h\) corresponds to the snapshot block height and \(u=t\) to month \(t\). Let \(b_i(u)\) denote the BTC balance of entity \(i\) at observation point \(u\). Aggregate BTC holdings are then
\[
B(u)=\sum_{a\in\mathcal{A}} B_a(u), \qquad 
B_a(u)=\sum_{i\in\mathcal{E}_a} b_i(u).
\]
Because many labeled entities hold zero BTC at the snapshot height, unconditional cross-sectional measures are dominated by zeros. We therefore restrict our analysis to entities with positive balances, as our study of Bitcoin (BTC) wealth concentration at the economic-agent level is, by definition, meaningful only for those entities that currently hold observable BTC.
\[
\mathcal{E}_a^{+}(u)=\{\,i\in\mathcal{E}_a : b_i(u)>0\,\},
\qquad
n_a^{+}(u)=|\mathcal{E}_a^{+}(u)|.
\]

\paragraph*{Top-$K$ Share Ratios, adapted from \cite{juodis_overview_2024}}
Let \(\mathrm{Top}_{a,k}(u)\) denote the set of top \(k\%\) positive-balance entities in class \(a\) at observation point \(u\), where \(k\in\{1,5,20\}\). The Top-$K$ share ratio is
\begin{equation}
\label{eq:topk}
T_{a,k}(u)=
\frac{\sum_{i\in \mathrm{Top}_{a,k}(u)} b_i(u)}
{\sum_{i\in \mathcal{E}_a^{+}(u)} b_i(u)}
\times 100\%.
\end{equation}
It measures the share of BTC held by the largest \(k\%\) of positive-balance entities within class \(a\). In implementation, the selected set always contains at least one entity.

\paragraph*{Gini Coefficient, adapted from \cite{sai_characterizing_2021}}
The Gini coefficient for positive-balance entities in class \(a\) at observation point \(u\) is
\begin{equation}
\label{eq:gini}
G_a(u)=\frac{1}{2[n_a^{+}(u)]^2\mu_a^{+}(u)}
\sum_{i=1}^{n_a^{+}(u)}\sum_{j=1}^{n_a^{+}(u)}|b_i(u)-b_j(u)|.
\end{equation}
Here, \(n_a^{+}(u)\) and \(\mu_a^{+}(u)\) denote the number and mean balance of positive-balance entities in class \(a\).

\paragraph*{Wealth Concentration Over Time}
Monthly series \(t\) of positive-entity counts \(\mathcal{E}_a^{+}\), \(G_a(t)\) is the Gini coefficient of class \(a\) over time \(t\), and \(T_{a,k}(t)\) are calculated for all agent-classes and overlaid with external event markers \(\{(\tau,\ell)\}\) where \(\tau\) is the date of the event marker \(\ell\) e.g., (2024-04, \quotes{Halving 2024}) shown in Fig.~\ref{fig:btc-wealth-concentration-2009-2025}, allowing observation of regime-dependent changes in wealth concentration and entity growth, and the emergence of persistent patterns of large-entity wealth concentration across agent-classes.

\begin{algorithm}[htbp]
\caption{Crypto-Microeconomic Observability for Economic Agents}
\label{alg:cmof}
\footnotesize
\begin{algorithmic}[1]
\Require Snapshot height \(h=905314\); Bitcoin node data; monthly grid \(t=t_0{:}t_1\); event set \(\{(\tau,\ell)\}\)
\State Extract and transform on-chain data using \textsc{BitSQL}.
\State Apply Multi-input clustering and tagging using BlockSci to form \(\mathcal{E}_a\), compute snapshot balances at \(h\), and reconstruct monthly balances \(b_i(t)\).
\State Compute snapshot descriptive statistics and snapshot inequality metrics by agent (or class), and plot the time-series analysis.
\ForAll{months \(t\in[t_0,t_1]\)}
  \ForAll{agents \(a\in\mathcal{A}\)}
    \State Construct \(\mathcal{E}_a^{+}(t)\) and compute positive-entity counts, \(B_a(t)\), \(G_a(t)\), and \(T_{a,k}(t)\) for \(k\in\{1,5,20\}\).
  \EndFor
\EndFor

\end{algorithmic}
\end{algorithm}

\section{Implementation and Evaluation}
\subsection{Experimental Setup and Dataset Overview}
Following Algorithm~\ref{alg:cmof}, the ETL pipeline uses a local Bitcoin node, \textsc{BitSQL} \cite{mun_bitsql_2022} for full-ledger extraction (block hight \texttt{905314}, at 2025-07-13 03:23:34 UTC) and \textsc{Watch Your Back-based} BlockSci for multi-input clustering and tagging \cite{gomez_watch_2022}. It ran on a Ubuntu (Intel Xeon CPU, 128 GB RAM, 6 TB NVMe SSD). These implementation components correspond to the first three stages of the observability method, namely ledger data extraction, entity-level clustering, and class attribution. The subsequent stages reconstruct account balances and compute concentration measures at the class level, thereby providing the empirical foundation for this letter’s Bitcoin's economic-agent classes analysis.

The empirical results indicate how BTC balances are concentrated across different economic agents in the Bitcoin ecosystem. Table~\ref{tab:dataset_overview} shows that the labeled dataset is highly sparse: only 8,450 of 223,558 entities have positive BTC balances (96.22\% are zero-balance). Service is the dominant class, with 71.98\% of entities and 75.15\% of observed BTC. Abuse accounts for just 3.53\% of entities but 24.26\% of BTC and has the highest mean balance (29.06 BTC), indicating much larger average holdings. From an economic perspective, the clustering under Service reflects aggregation around service and intermediary entities, whereas the outsized Abuse share shows that illicitly designated entities form a financially significant stock of BTC that is sensitive to enforcement actions. In contrast, Individuals and Malware have many entities but negligible BTC shares, so prevalence does not necessarily reflect wealth concentration.

\begin{table}[t]
\caption{Dataset Overview by Economic Agent}
\label{tab:dataset_overview}
\centering
\footnotesize
\setlength{\tabcolsep}{3pt}%
\renewcommand{\arraystretch}{0.95}%
\begin{tabular}{@{}lrrrrrrr@{}}
\toprule
\shortstack{\textbf{Agent}\\\textbf{(Class)}} & \shortstack{\textbf{Entities}\\\textbf{(Ents.)}}  & \shortstack{\textbf{Pos.}\\\textbf{Ents.}} & \shortstack{\textbf{\% 0-Bal.}\\\textbf{Ents.}} & \shortstack{\textbf{\%}\\\textbf{Ents.}} & \shortstack{\textbf{Total}\\\textbf{BTC}} & \shortstack{\textbf{\%}\\\textbf{Supply}} & \shortstack{\textbf{Mean}\\\textbf{BTC}} \\
\midrule
Service & 160,923 & 917 & 99.43 & 71.98 & 709,946.04 & 75.15 & 4.41 \\
Abuse & 7,888 & 1,003 & 87.28 & 3.53 & 229,227.65 & 24.26 & 29.06 \\
Malware & 13,133 & 2,611 & 80.12 & 5.87 & 170.02 & 0.02 & 0.01 \\
Individuals & 41,600 & 3,907 & 90.61 & 18.61 & 5,408.54 & 0.57 & 0.13 \\
Benign & 14 & 12 & 14.29 & 0.01 & 3.43 & 0.00 & 0.24 \\
\midrule
\textbf{Total} & \textbf{223,558} & \textbf{8,450} & \textbf{96.22} & \textbf{100} & \textbf{944,755.69} & \textbf{100} & \textbf{4.23} \\
\bottomrule
\end{tabular}
\end{table}

\subsection{Descriptive Statistics of Bitcoin Wealth by Economic Agent}
Table~\ref{tab:descriptive_stats} indicates that BTC balances are extremely right-skewed and heavy-tailed for all labeled economic-agent classes. More importantly, beyond this expected skewness in Bitcoin balances, the extent and economic meaning of concentration vary systematically across classes and masks different economic roles. Service and Abuse have by far the largest averages and dispersion (means of 774.21 BTC and 228.54 BTC, maxima of 248,597.57 BTC and 79,957.27 BTC), while Malware, Individuals, and Benign hold much smaller averages. Very low medians in all classes indicate that most positive-balance entities hold negligible BTC. Large mean–median gaps and very high skewness and kurtosis confirm that the balances are concentrated in a few large entities, especially individuals, which show the highest skewness (62.24) and kurtosis (3,883.87), reflecting the most extreme tail behavior and strongest large-entity wealth concentration.

\begin{table}[t]
\caption{Descriptive Statistics by Economic Agent}
\label{tab:descriptive_stats}
\centering
\footnotesize
\setlength{\tabcolsep}{2.2pt}%
\renewcommand{\arraystretch}{0.95}%
\begin{tabular}{@{}lrrrrr@{}}
\toprule
\textbf{Statistic} & \textbf{Service} & \textbf{Abuse} & \textbf{Malware} & \textbf{Individuals} & \textbf{Benign} \\
\midrule
Mean (BTC) & 774.21 & 228.54 & 0.07 & 1.38 & 0.29 \\
Median (BTC) & 0.01 & 0.00 & 0.05 & $<10^{-4}$ & 0.02 \\
Std Dev (BTC) & 10,864.81 & 3,737.12 & 0.57 & 78.95 & 0.43 \\
Skewness & 17.70 & 18.64 & 23.14 & 62.24 & 1.19 \\
Kurtosis & 349.13 & 359.64 & 548.20 & 3,883.87 & 2.77 \\
Min (BTC) & $<10^{-7}$ & $<10^{-5}$ & $<10^{-7}$ & $<10^{-7}$ & $2.03\times10^{-4}$ \\
Max (BTC) & 248,597.57 & 79,957.27 & 15.41 & 4,928.73 & 1.25 \\
\bottomrule
\end{tabular}
\end{table}
\subsection{Inequality Metrics by Economic Agent}

Table~\ref{tab:inequality_metrics} shows severe within-class wealth concentration in Service, Abuse, and Individuals, with near-unit Gini coefficients (0.9935, 0.9962, 0.9993), Top-1\% shares above 95\%, Top-5\% and Top-20\% shares almost fully concentrated. Malware and Benign are less concentrated, with lower Gini coefficients and more moderate Top-$K$ shares. Overall, concentration differs by economic agent's role: in the Service, Abuse, and Individual categories, a small number of large-entities dominate, suggesting that concentration arises from mechanisms specific to each agent class rather than from a single, uniform pattern across the entire Bitcoin ecosystem.

\begin{table}[t]
\caption{Inequality Metrics by Economic Agent}
\label{tab:inequality_metrics}
\centering
\footnotesize
\setlength{\tabcolsep}{3pt}%
\renewcommand{\arraystretch}{0.95}%
\begin{tabular}{@{}lrrrrr@{}}
\toprule
\textbf{Metric} & \textbf{Service} & \textbf{Abuse} & \textbf{Malware} & \textbf{Individuals} & \textbf{Benign} \\
\midrule
Gini Coefficient & 0.9935 & 0.9962 & 0.5380 & 0.9993 & 0.7239 \\
Top-1\% Share & 95.8030 & 99.5545 & 42.9816 & 99.5952 & 36.3414 \\
Top-5\% Share & 99.5269 & 99.9658 & 46.2693 & 99.9829 & 36.3414 \\
Top-20\% Share & 99.9851 & 99.9990 & 57.8145 & 99.9994 & 64.5625 \\
\bottomrule
\end{tabular}
\end{table}

\subsection{Concentration of Bitcoin Wealth Among Economic Agents Over Time}
\begin{figure}[htbp]
  \centering
      \includegraphics[
    width=\linewidth,
    height=\textheight,
    keepaspectratio
  ]{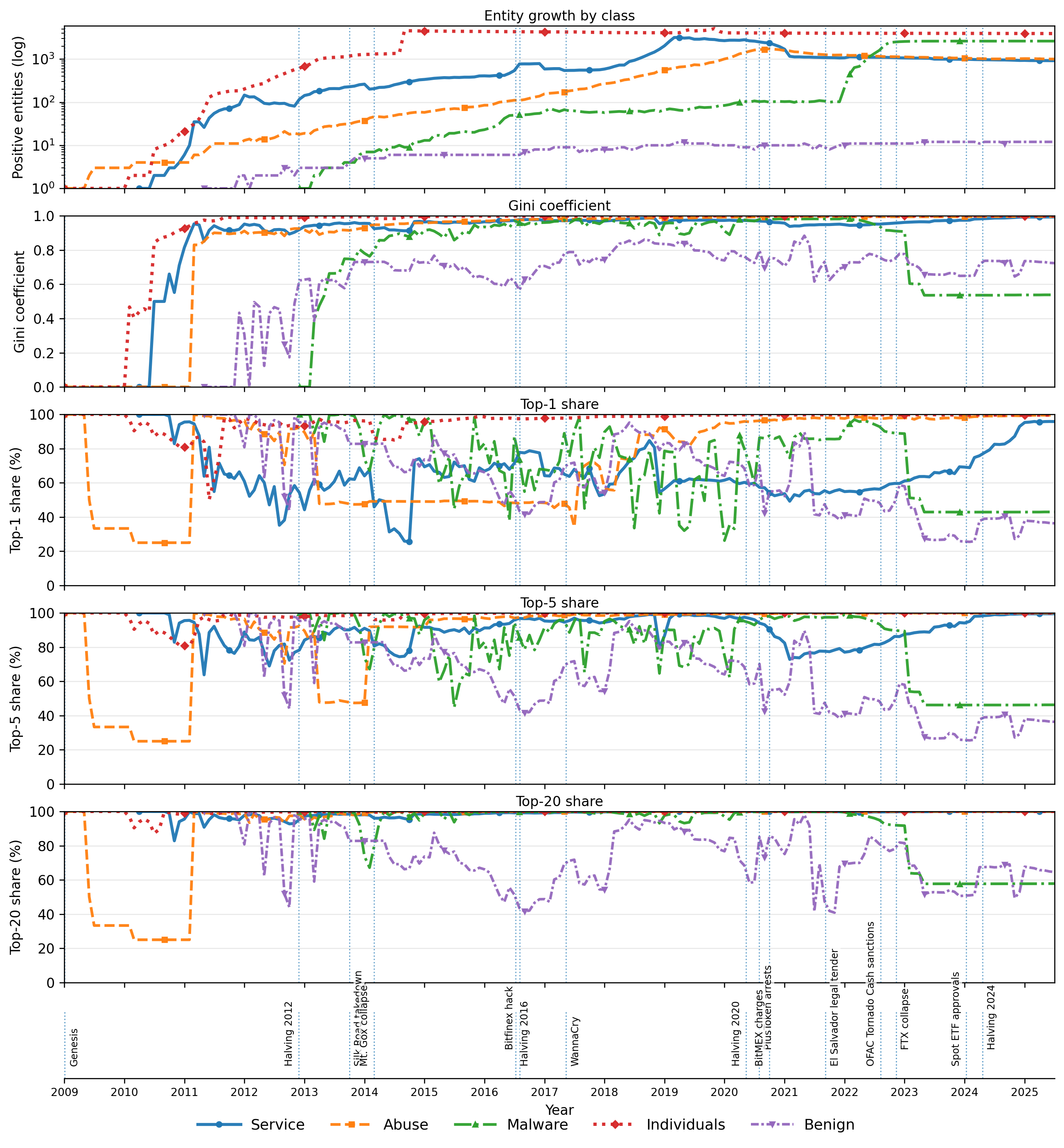}
  \caption{Changes in Bitcoin wealth concentration (Top-1 / Top-5 / Top-20 shares, 2009–2025), highlighted with dotted markers to indicate significant market events.}
  \label{fig:btc-wealth-concentration-2009-2025}
\end{figure}

Fig.~\ref{fig:btc-wealth-concentration-2009-2025} presents the evolution of Bitcoin agent-class level wealth concentration between 2009 and 2025 for the Top-1, Top-5, and Top-20 shares, together with major \quotes{crypto-financial events}, including exchange collapses, enforcement actions, halving cycles, and regulatory shifts. These events include the Genesis block \cite{mun_bitsql_2022}, the Halving Episodes \cite{LASHKARIPOUR2024106198}, the Silk Road, the Mt. Gox collapse, and the Bitfinex hack \cite{FENG201863}, WannaCry Ransomware \cite{Oosthoek2023}, BitMEX charges \cite{cftc2020bitmex}, PlusToken arrest \cite{plustokenarrest2020}, El Salvador legal tender \cite{ElSalvadortendor}, OFAC Tornado Cash sanctions \cite{brownworth2024tornado}, FTX collapse \cite{YOUSAF2023103661}, and spot Bitcoin ETFs approval by the U.S. SEC \cite{LIU2024106150}. This shows that concentration remains consistently high for Service, Abuse, and Individuals, whereas it is lower and more variable for Malware and Benign. Also, it reveals whether high concentration persisted across market regimes and whether it remained anchored in similar agent classes. The series shows repeated waves of concentration and partial dispersion, highlighting the persistent role of large-entities, with BTC wealth largely confined to a small subset of entities in most agent-classes.

\subsection{Discussion}
The microeconomic analysis shows strong asymmetry in Bitcoin wealth across agent classes. Service-labeled entities collectively hold the majority of the observable BTC supply, while Abuse holds a disproportionately large share relative to its size, so wealth concentration cannot be explained by entity count alone. Within classes, balances are highly skewed: most entities hold negligible BTC, and a small subset holds nearly all observed funds. Longitudinal patterns indicate this structure is persistent, especially for Service, Abuse, and Individuals, rather than tied to short-lived market episodes.

Also, the agent-class decomposition directly connects Bitcoin concentration to the financial sector through market structure, in contrast to prior work that links concentrated cryptocurrency ownership mainly to concerns such as market manipulation and Bitcoin price behavior. In this letter, Service concentration refers to the clustering of balances around custodians, exchanges, payment processors, and other transaction-focused intermediaries. Abuse and Malware concentration isolates balances associated with enforcement-relevant activities and illicit transfers, while Individual concentration captures exposure to large private holders. These class-specific patterns identify where potential crypto-financial risk channels are located within the labeled, positive-balance sample.

\section{Conclusion and Future Work}

In this letter, Crypto-Microeconomic observability is applied to measure and analyze the distribution of Bitcoin-denominated wealth across five labels: Service, Abuse, Malware, Individual, and Benign. The findings show structurally persistent concentration, providing evidence of a large-entities BTC concentration: Service holds 75.15\% of the observed BTC supply, while within-class inequality reaches near-maximal levels, including a Gini coefficient of 0.9993 for Individuals and a Top-1\% share of 99.5545\% for Abuse. The findings indicate that Bitcoin’s overall concentration can be decomposed into distinct, economically relevant agent classes, rather than viewed solely as one network-wide measure. This viewpoint also highlights how intermediation, enforcement-sensitive activity, illicit-payment flows, and large-entity concentration networks emerge within the Bitcoin ecosystem. Future work will extend the methodology with improved entity-resolution and attribution methods to better identify, cluster, and track agent's micro-level wealth inequality.

\printbibliography

\section*{Statements and Declarations}

\subsection*{Funding}

The authors declare that they did not receive any financial support, including funds or grants, for the preparation of this manuscript.

\subsection*{Competing Interests}

The authors declare that they have no financial or non-financial conflicts of interest relevant to the content of this work.

\subsection*{Author Contributions}

Syed Azhar Hussain: writing - original draft, writing - review and editing, methodology, conceptualization, data acquisition, and data analysis. Mubashir Husain Rehmani: supervision, writing – review and editing. Kashif Ahmad: supervision, writing - review and editing. 

\subsection*{Data Availability}

The authors obtained the data for this study from publicly available Bitcoin blockchain records, using the extraction, clustering, classification, and processing methods described in the Methodology section. The resulting datasets supporting this work can be requested from the corresponding author.

\subsection*{Ethics Approval}

Not applicable. This study used only publicly available Bitcoin blockchain data and did not involve human participants, biological specimens, or animals.

\end{document}